\documentclass[aps,prb,twocolumn,showpacs,groupedaddress]{revtex4-1}

\usepackage{graphicx}
\usepackage{dcolumn}
\usepackage{bm}
\usepackage{amsmath}
\usepackage{natbib}
\usepackage{lineno}
 \usepackage{array}

\begin{document}

\preprint{APS/123-QED}

\title{Irreversible magnetization switching using surface acoustic waves}


\author{L. Thevenard$^{1}$, C. Gourdon$^{1}$, J.-Y. Duquesne$^{1}$, E. Peronne$^{1}$,  H. J. von Bardeleben$^{1}$, H. Jaffres$^{2}$,  S. Ruttala$^{2}$ and A. Lema\^itre$^{3}$}

 
\affiliation{
$^1$ Institut des Nanosciences de Paris, Universit\'{e} Pierre et Marie Curie, CNRS, UMR7588, 4 place Jussieu,75252 Paris, France\\
$^2$ Unit\'{e} Mixte de Physique CNRS/Thales and Universit\'{e} Paris Sud 11, Route D\'{e}partementale 128, 91767 Palaiseau, France\\
$^3$ Laboratoire de Photonique et Nanostructures, CNRS, UPR 20, Route de Nozay, Marcoussis, 91460, France \\}

\date{\today}

\label{sec:Abstract}

\begin{abstract}


An analytical and numerical approach is developped to pinpoint the optimal experimental conditions to irreversibly switch  magnetization using surface acoustic waves (SAWs). The layers are magnetized perpendicular to the plane and two switching mechanisms are considered. In precessional switching, a small in-plane field initially tilts the magnetization and the passage of the SAW modifies the magnetic anisotropy parameters through inverse magneto-striction, which triggers precession, and eventually reversal. Using the micromagnetic parameters of a fully characterized layer of the magnetic semiconductor (Ga,Mn)(As,P), we then show that there is a large window of accessible experimental conditions (SAW amplitude/wave-vector, field amplitude/orientation) allowing irreversible switching. As this is a resonant process, the influence of the detuning of the SAW frequency to the magnetic system's eigenfrequency is also explored. Finally, another - non-resonant - switching mechanism is briefly contemplated, and found to be applicable to (Ga,Mn)(As,P): SAW-assisted domain nucleation. In this case, a small perpendicular field is applied opposite the initial magnetization and the passage of the SAW  lowers the domain nucleation barrier.

\end{abstract}


\maketitle

\section{Introduction}

In a large number of ferromagnets, the coupling between strain and magnetization originates from the spin-orbit interaction, and was shown early on to be maximum when elastic and magnetic resonance (precession) frequencies match\cite{bommel59}. This effect has been revisited in the light of spintronics applications in the past few years with compelling dynamic experiments in both magnetic semiconductors\cite{Scherbakov2010,Bombeck2012} and metals\cite{Weiler2011}. A first approach relies on the generation of picosecond acoustic pulses (longitudinal or transverse phonons). When coupled to the layer's magnons,  magnetization precession may be triggered\cite{Bombeck2012}, but it remains a fairly inefficient mechanism as the strain spectrum peaks quite high (20-30~GHz\cite{Thevenard2010a}) above typical precession frequencies (0.5-10~GHz). Switching of a perpendicularly magnetized (Ga,Mn)(As,P) structure has recently been demonstrated using this technique\cite{Casiraghi2011}, but the effect was shown to originate from incoherent phonons (heat waves), and not from a magneto-strictive effect due to the high frequency coherent  phonons produced.  Another route consists in generating strain through lower frequency ($<$2 GHz) surface acoustic waves (SAWs). On in-plane magnetized systems, SAWs have  been used to drive ferromagnetic resonance in thin Ni films\cite{Weiler2011}, or periodically switch magnetization between hard and easy axes in Co bars\cite{Davis2010}. Recent theoretical work has focused on the switching of in-plane Terfenol nanomagnets subjected to stress\cite{Roy2011a,Fashami2011}, but no experimental or theoretical work has been shown on perpendicularly magnetized systems. These materials are for instance particularly relevant to high density magnetic information storage technologies. We believe SAWs offer two main advantages for magnetization reversal compared to picosecond acoustics: their relatively low frequencies, easily matched to precession frequencies, and the narrow bandwidth of the generated acoustic wave (a few MHz), as opposed to the broad-band spectrum in the former technique.

In this work, we wish to address theoretically the irreversible magnetization reversal  in perpendicularly magnetized layers using  surface acoustic waves, and  under realistic experimental conditions on a test system consisting in thin (Ga,Mn)(As,P) layers, a magneto-strictive dilute magnetic semiconductor. Two possible mechanisms are considered, both relying on the transient modification of the magnetic anisotropy by the SAW. In precessional switching, the magnetization is pulled away from equilibrium by an in-plane field, and the SAW triggers a large angle precession of the magnetization which may end up in a full reversal. In SAW-assisted domain nucleation, a small perpendicular field is applied opposite the initial magnetization, and the SAW is used to locally lower the domain wall (DW)  energy, and thus initiate domain nucleation, leading to a full reversal.

\section{Description of the system}
\label{sec:DescriptionOfTheSystem}

\subsubsection{Generation of SAWs}

\begin{figure}
	\centering
		\includegraphics[width=0.3\textwidth]{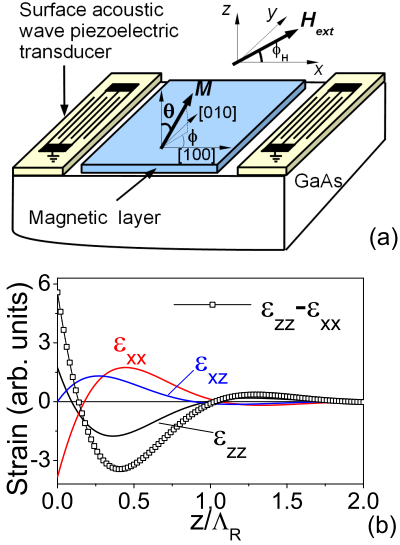}
	\caption{(a) Set-up geometry for a SAW propagating along [100], and coordinates. (b) Depth dependence of the amplitude of the Rayleigh wave components ($f$=1~GHz) plotted using Annex B equations.}
	\label{fig:StrainWave}
\end{figure}

SAWs are excited and detected by interdigital transducers (IDTs) on a piezoelectric layer\cite{Duquesne2012,Sanada2011} deposited on a magnetic thin film (Fig. \ref{fig:StrainWave}a). We will for now limit ourselves to the case of a Rayleigh wave propagating along the [100] axis of a cubic crystal. The case of a wave propagating along [110] will be discussed in Section IV.2. The only finite  propagating strain wave components are  then $\varepsilon_{xx}(x,z,t) $,  $\varepsilon_{zz}(x,z,t)$, and $\varepsilon_{xz}(x,z,t)$ (details in  Annex B, axes defined in Fig. \ref{fig:StrainWave}a). Their wavelength is given directly by the IDT period $\Lambda_{R} \approx$ 3-5~$\mu$m for $f$=0.5-1 GHz, and their dispersion-free velocity by the elastic constants of the material, $V_{R}$=2711 m.s$^{-1}$, with  $\Lambda_{R}$=$\frac{V_{R}}{f}$. Their depth-dependence is plotted in Fig. \ref{fig:StrainWave}b. Two hypotheses may then be made if the magnetic layer is taken much thinner ($<$50 nm) than $\Lambda_{R}$: (i) the $\varepsilon_{xz}$ component can be neglected, as its amplitude remains weak close to the surface, and (ii) the strain field can be considered constant along $z$. We will therefore take $z$=0 in the expressions of $\varepsilon_{zz}$, $\varepsilon_{xx}$ and $\varepsilon_{xz}$=0.  Finally, the RF power passing through the combs is small enough (10~mW) to neglect any resulting heating of the sample.

\subsubsection{Magnetic system}

The time-dependent dynamics of the magnetization are described by the Landau-Lifshitz-Gilbert (LLG) equation:

     \begin{align}
     \label{eq:LLGa}
\frac{\partial\vec{M}}{\partial t} &= -\gamma \vec{M} \times \mu_{0}\vec{H}_{eff}+\frac{\alpha}{M_{s}} \vec{M} \times \frac{\partial{\vec{M}}}{\partial t} \\
     \label{eq:LLGb}
\mu_{0} \vec{H}_{eff}&= -\vec{\nabla}_{M}F(\vec{M})  
      \end{align}

where $\vec{M}[\theta(x,t),\phi(x,t)]$ is the magnetization expressed in polar coordinates with $M_{s}$ its norm (taken to be constant) and  $\gamma>0$  the gyromagnetic ratio. $ \vec{H}_{eff}$ is the effective field, i.e the sum of a magneto-crystalline anisotropy term, a shape anisotropy term, and finally the Zeeman contribution from the externally applied field. In this work (except in Section V), the exchange energy contribution will be neglected and we will effectively be looking at the behavior of a single macrospin.

 Following Linnik \textit{et. al}\cite{Linnik2011}, a normalized free energy density $F_{M}$=$F/M_{s}$ is defined, where $F$ is a very general form of the free energy density of a cubic ferromagnetic layer
distorted by strain:

      \begin{align}
      \label{eq:Fm}
      \begin{split}
F_{M}(\theta,\phi)  = (A_{2\varepsilon}-2A_{4\varepsilon}) \varepsilon (x,t)\cos^{2} \theta +\\
 (B_{c}+2A_{4\varepsilon} \varepsilon (x,t))\cos^{4}\theta+\\
 \frac{1}{4}\sin^{4}\theta(B_{c}-A_{4\varepsilon} \varepsilon (x,t))(3+\cos 4\phi)+\\
  \frac{\mu_{0}M_{s}}{2}\cos^{2} \theta  + \frac{1}{2}A_{2xy}\varepsilon_{xy}\sin^{2}\theta \sin 2\phi-\\
  [\sin \theta (\mu_{0}H_{x}\cos \phi +  \mu_{0}H_{y}\sin \phi) +\mu_{0}H_{z} \cos \theta]  
      \end{split} 
       \end{align}

$\vec{H}_{ext}$=($H_{x}$,$H_{y}$,$H_{z}$) is the externally applied field and $\theta(x,t),\phi(x,t)$ is abbreviated into $\theta$,$\phi$. $B_{c}$ is the cubic anisotropy field, and $A_{4\varepsilon}$, $A_{2\varepsilon}$, $A_{2xy}$ are the magnetoelastic coefficients. The dependence of the magnetic anisotropy on strain is given through the terms $\varepsilon_{xy}$ (static shear strain, details in Annex A), and $ \varepsilon(x,t)$= $\Delta \varepsilon_{0}+\delta \varepsilon (x,t)$.  $\delta \varepsilon (x,t)$=$\delta \varepsilon_{zz} (x,t)-\delta \varepsilon_{xx} (x,t)$ is the strain generated by the SAW. The Rayleigh wave propagating in a cubic material along $\vec{q}$//[100] has been calculated analytically, and found to be quite different from the one used for isotropic materials [Ref JYQ] (details in Annex B). The resultant strain difference is given by\cite{note1_acoustic_2012}:

     \begin{align}  
\label{eq:epsSAWc} 
\delta \varepsilon (x,t) & =  \varepsilon_{max}\cos(\omega t-qx)
        \end{align}

where $\varepsilon_{max}$ is the SAW amplitude\cite{note1_acoustic_2012}, and $q$ its wave-vector, with $q$=$\omega/V_{R}$. $\Delta \varepsilon_{0}$=$\varepsilon_{zz,0}-\varepsilon_{xx,0}$ is the difference between the static out-of-plane and in-plane strain components, related by the elastic modules C$_{11}$, C$_{12}$ of GaAs\cite{note2_acoustic2012} through $\varepsilon_{zz,0}$=-2$\frac{C_{12}}{C_{11}}\varepsilon_{xx,0}$. 

\setcounter{subsubsection}{0}

\section{Analytical solution: small angle precession}
\label{sec:AnalyticalSolutionSmallAnglePrecession}

\subsubsection{Principles of precessional switching}

 In so-called precessional switching, the perpendicular magnetization $\vec{M}$ is first tilted towards the layer by an in-plane magnetic field. A short perturbation (e.g an optical\cite{Kimel2004b,DeJong2012}, acoustic\cite{Bombeck2012}, or ultra-fast magnetic\cite{Gerrits2002} or electric\cite{Balestriere2011a} field pulse) then modifies the micromagnetic parameters enough to change the effective field seen by the magnetization, and send it precessing. If the precession amplitude is sufficiently large, the magnetization can switch to another potential valley, where it will remain if the perturbation lasts an odd multiple of half the precession period\cite{Gerrits2002}, or if damping eventually prevents $\vec{M}$ from oscillating between the two minima\cite{} ("ringing" phenomenon). This mechanism has for instance being suggested  for  micro-wave assisted switching  at a head field significantly below the medium coercivity\cite{Zhu2008} or for subnanosecond spin torque switching in magnetic tunnel junctions\cite{Rowlands2011}.  \\

  The arrival of the SAW on the magnetic layer leads to a modification of the magneto-strictive  anisotropy terms, and thereby of the effective field  $\vec{H}_{eff}$. The time response of the magnetization is assumed  short (around 100~ps) on the SAW's  time-scale, as evidenced by recent pump-probe experiments\cite{Scherbakov2010}. Eq. \eqref{eq:LLGa} then shows  that this  triggers magnetization precession as long as the torque $\vec{M}\times \mu_{0}\vec{H}_{eff}$ remains non-zero and the damping has not aligned $\vec{M}$ back along the applied field. Two computational approaches were  then followed. Firstly,  the conditions leading to magnetization precession were established by assuming small changes in magnetization direction, $\delta \theta$, $\delta \phi$, in order to solve this equation analytically. Secondly, in view of establishing the experimental conditions leading to irreversible precessional switching of a (Ga,Mn)(As,P) layer, the LLG equation was   solved numerically, and a switching diagram established. In this work, perpendicularly magnetized layers were considered. This is often a problematic configuration, since the energy barriers are high for a full $\pi$ reversal of the magnetization.

\subsubsection{General solution}
\label{sec:GeneralSolution}

    In this first approach, perturbations are small, leading to small changes in the magnetization direction around its equilibrium position $\vec{M_{0}}[\theta_{0},\phi_{0}]$. Provided the magnetic anisotropy and applied fields are such that $\theta_{0}\neq$0, Eq. \eqref{eq:LLGa}  can be linearized  into:

\begin{align}
\label{eq:SAPa}
	-\dot{\delta \theta} &=\frac{\gamma}{\sin \theta_{0}} [F_{\phi \phi}\delta \phi+F_{\phi \theta}\delta \theta+F_{\phi \varepsilon}\delta \varepsilon]+\alpha \dot{\delta \phi}\sin \theta_{0} \\
\label{eq:SAPb}
	\dot{\delta \phi} &=\frac{\gamma}{\sin \theta_{0} } [F_{\theta \theta}\delta \theta+F_{\theta \phi}\delta \phi+F_{\theta \varepsilon}\delta \varepsilon]+\frac{\alpha}{\sin \theta_{0}} \dot{\delta \theta}
\end{align}

The terms $F_{ij}$ stand for $\frac{\partial F_{M}}{\partial i \partial j}$, and the dot denotes the time derivative. 
In the following, the magnetization precession amplitude $\delta \theta(x,t)$ will be calculated in $x$=0, but can easily be obtained at any distance $x$ from the comb by computing $\delta \theta(t-x/V_{R}$).

The eigen-frequency of the system in the absence of acoustic wave is first determined by assuming harmonic solutions for the angle deviations: $\delta \theta=\delta \theta_{0}e^{i\Omega_{P}t}$, $\delta \phi=\delta \phi_{0}e^{i\Omega_{P}t}$. Expressing the determinant of the corresponding coupled equations system (5,6) then yields the complex precession frequency $\Omega_{P}$ of the magnetization in the presence of a finite damping term where  we define $\Omega_{P}$=$\omega_{P}$+$i\chi$:

\begin{align}
\label{eq:FreqPrec}
 \omega_{P}&=\frac{1}{\sqrt{1+\alpha^{2}}}\sqrt{\omega^{2}_{0}-\frac{\alpha^{2}\gamma^{2}H^{2}_{\alpha}}{4(1+\alpha^{2})}}\\
	\chi&=\frac{\alpha \gamma H_{\alpha}}{2(1+\alpha^{2})}\\	
	\omega_{0}&=\frac{\gamma}{sin \theta_{0}}\sqrt{F_{\theta \theta}F_{\phi \phi}-F^{2}_{\theta \phi}}	
\end{align}

We have further defined  an effective field $H_{\alpha}$=$F_{\theta \theta}+F_{\phi \phi}/sin^{2}\theta_{0}$, and  the precession rate $\omega_{0}$ in the absence of damping. Assuming the SAW  arrives at an instant $t$=0, the variation-of-parameters method then yields the amplitude of the magnetization precession $\delta\theta(t)$ as a function of the exciting SAW frequency $\omega$ and amplitude $\varepsilon_{max}$, the precession frequency $\omega_{P}$ and the  damping $\alpha$: 

\begin{widetext} 
      \begin{align} \label{eq:deltatheta0}
\delta\theta (t)= \frac {\varepsilon_{max}\Omega_{\theta}}{(1 + \alpha ^{2})\sqrt{(\omega^{2} - \omega^{2}_{res})^{2}+ \Gamma ^{4}}}\left [f(\omega,\beta)\cos(\omega t+\eta) - \frac {\omega_{0} e^{-\chi t}}{\sqrt{1 + \alpha ^{2}}}\cos(\omega_{P} t +\xi) \right],
		 \end{align}		
\end{widetext}

where $\xi$ and $\eta$ are two phase shifts that depend on $\omega$ and the material's parameters. $f(\omega,\beta)$, $\beta$ and $\Omega_{\theta}$ are  defined in  Annex D. We further define: 

  \begin{align}\label{eq:GammaOmega}
\omega^{2}_{res}&=\omega^{2}_{P}-\chi^{2}\\
\Gamma&=\sqrt{2\omega_{P}\chi}
  \end{align}

$\omega_{res}$ is the resonance frequency of the  system and $\Gamma$ is related to the resonance broadening. This very general expression of the precession amplitude highlights two physical behaviors. The first one is that, as expected intuitively, the precession consists of a forced term oscillating at the excitation frequency $\omega$, and a damped term at the eigenfrequency of the magnetic system  $\omega_{P}$. The second one if that the excitation frequency giving the largest amplitude is not exactly  $\omega_{P}$, but a  slightly lower value,  $\omega_{res}$, which is a modified resonance frequency of the damped  system in the small perturbation regime. Finally, a broadening term $\Gamma$ allows the precession amplitude not to diverge at resonance. 

The amplitude of the precession is linear in $\varepsilon_{max}$ and  in $\Omega_{\theta}$ which depends non-trivially  on the magneto-strictive coefficients, the damping, and the applied field through the value of $\theta_{0},\phi_{0}$ (details in Annex D). We will see below that its expression can however be greatly simplified in some limiting cases.

\subsubsection{Application to thin (Ga,Mn)(As,P) layers}
\label{sec:AnApplicationToThinGaMnAsLayers}

The dilute magnetic semiconductor (Ga,Mn)(As,P) is a good test-bench material to investigate fast acoustics-induced magnetization switching. The carrier-mediated nature of its ferromagnetic phase results in a strong dependence of the magnetic anisotropy on the strain state of the layer, through the band splitting it induces on the light- and heavy-hole bands\cite{Dietl2001}. For instance, Glunk et. al\cite{Glunk2009} have shown that the perpendicular uniaxial anisotropy term in (Ga,Mn)(As,P) is proportional  to both the out-of-plane strain coefficient $\varepsilon_{zz}$ and the hole concentration $p$. Moreover, contrary to metals,  typical precession frequencies of (Ga,Mn)(As,P) can be fairly low, of the order of the GHz in small magnetic fields\cite{Thevenard2010a},  which  allows good matching to the acoustic wave frequencies provided by SAWs. 
 Finally, the damping parameter can be rather high in this material ($\alpha$=0.1-0.3\cite{Gourdon2007,Nemec12}) compared to metals ($\alpha$=0.01 in Ni$_{80}$Fe$_{20}$), which will limit  ringing effects preventing irreversible switching. Whereas in metals, precessional switching is mainly governed by the precession of the magnetization around the demagnetizing field, in (Ga,Mn)(As,P) this process will  be driven by the magneto-crystalline anisotropy since its  magnetization at saturation is weak.

 While Eq. \eqref{eq:Fm} conveniently  highlights the magneto-strictive terms, the following form of energy is more commonly used\cite{Farle1998} to determine experimentally the anisotropy coefficients in (Ga,Mn)(As,P) (details in Annex A for the correspondence between both energy forms):

\begin{equation}
\label{eq:Gm}
\begin{split}
   F_{M}(\theta,\phi)=-B_{2\bot}\cos^{2} \theta  -\frac{1}{2}B_{4\bot}\cos^{4}\theta -\\
   \frac{1}{8}B_{4//}\sin^{4}\theta(3+\cos 4\phi)-B_{2//}\sin^{2}\theta \sin^{2}(\phi -\frac{\pi}{4})+\\
   \frac{\mu_{0}M_{s}}{2}\cos^{2}\theta-[\sin \theta (\mu_{0}H_{x}\cos \phi +  \mu_{0}H_{y}\sin \phi) + \mu_{0}H_{z} \cos \theta]  
\end{split}
\end{equation}

The magnetic anisotropy is largely dominated by the uniaxial term $B_{2\bot}$, followed by the cubic terms $B_{4\bot}$ and $B_{4//}$ which result from the tetragonal distortion of the lattice as the magnetic layer grows strained upon its substrate. A linear dependence of the uniaxial anisotropy on strain has indeed been found experimentally using various techniques\cite{Masmanidis2005,Glunk2009,Cubukcu2010a}. The in-plane uniaxial term $B_{2//}$ is weakest and corresponds to a minor anisotropy between [110] and [1$\overline{1}$0] axes (details in Annex A).

 To estimate quantitatively the amplitude of the precession, a sample of relatively small perpendicular anisotropy is chosen in order to have GHz or sub-GHz precession frequencies, adapted to SAWs excited by micron-wide IDTs. An existing 50~nm thick sample, with a static strain  $\varepsilon_{zz,0}$=-0.05$\%$ and $x_{Mn}\approx 7\%$ is considered. This small lattice mismatch, yielding a moderate magnetic anistropy, is obtained by co-doping the (Ga,Mn)(As,P) layer  with Phosphorus ($y_{P}\approx 4\%$) as described in Ref. \onlinecite{lemaitre08}. At 95~K, $M_{s}$~=~9~kAm$^{-1}$ and ferromagnetic resonance spectroscopy yields: B$_{2\bot}$~=~22.5~mT, B$_{4\bot}$~=~-2.3~mT, B$_{4//}$~=~2.3~mT and B$_{2//}$~=~-1.2~mT. The  damping will be taken as $\alpha$=0.1, but note that this term has been shown to vary between 0.001 and 0.3 with magnetic and electric doping, as well as whether one measures the extrinsic damping or an intrinsic Gilbert damping\cite{Khazen2008a, Nemec12,Gourdon2007}.

Let us first put some numbers on the relevant frequencies ($f_{k}$=$\omega_{k}/2\pi$). Under an  in-plane magnetic field of 2~mT ($\theta_{0}$=3$^{\circ}$), Eqs. (7,9) yield: $f_{0}$=1.017~GHz, $f_{P}$=1.007~GHz, $f_{res}$=1.002~GHz. The decrease of the resonance frequency due to the inclusion of damping is therefore relatively small, a mere 1.5$\%$. The broadening is rather average: 90~MHz (full-width at half maximum). Finally, the exponential damping of the precession occurs on a time-scale of $1/\chi$=1.6~ns.

In order to isolate the relevant parameters to obtain a large angle precession, $\Omega_{\theta}$ and $f(\omega,\beta)$ may be simplified provided the explicit energy density of (Ga,Mn)(As,P) (Eq. \eqref{eq:Gm}) is used and a few hypotheses are made. Since in general $\beta<1$ (see Annex D), we develop $f(\omega,\beta)\approx \omega_{P}+\chi\beta$. We  also use: $A_{4\varepsilon}<<A_{2\varepsilon}$\cite{note1_acoustic_2012}, $\alpha<<1$, and consider that $\phi_{0}$ closely follows the applied field direction $\phi_{H}$, so that $\mu_{0}H_{x}\cos\phi_{0}+\mu_{0}H_{y}\sin\phi_{0}\approx \mu_{0}H_{ext}$. The precession amplitude at resonance can then be simplified into:

\begin{widetext} 
\begin{equation}
 \label{eq:OmegaTheta0GaMnAs}
|\delta \theta|_{max} \approx \varepsilon_{max}\frac{\Omega_{\theta,0}(\omega_{P}+\chi\beta)}{\Gamma^{2}}\hspace{5 mm}  ,\hspace{5 mm} 
\Omega_{\theta,0} \approx \frac{8 \gamma^2}{\omega_{0}}A_{2\varepsilon} \cos \theta_{0}(B_{4//}\sin^{3}\theta_{0}\cos 4\phi_{0}-B_{2//}\sin \theta_{0}\sin 2\phi_{0}+\mu_{0}H_{ext}/2)
\end{equation}
\end{widetext} 

\begin{figure}
	\centering
		\includegraphics[width=0.5\textwidth]{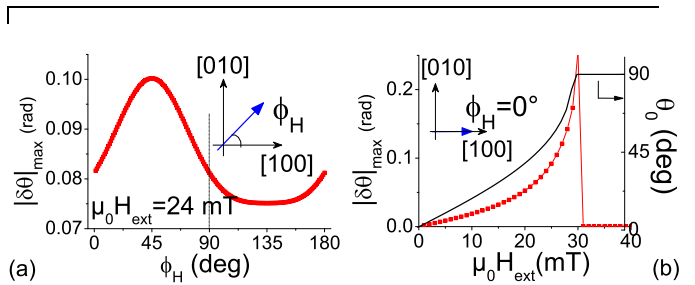}
	\caption{Conditions maximizing the precession amplitude in a perpendicularly magnetized (Ga,Mn)(As,P) layer  for $\varepsilon_{max}$=10$^{-5}$. (a) At fixed field amplitude, variation of the precession amplitude as a function of the field angle  $\phi_{H}$. (b) At fixed field angle $\phi_{H}$=0$^{\circ}$, variation of the precession amplitude as a function of the field amplitude  $\mu_{0}H_{ext}$ (red symbols) compared to the tilt $\theta_{0}$ (solid black line) before the arrival of the SAW.}
	\label{fig:SmallAnglePrecession}
\end{figure}

In Eq. \eqref{eq:OmegaTheta0GaMnAs}, $A_{2\varepsilon}$ is roughly proportionnal to the uniaxial anisotropy term $B_{2\bot}$ (see Annex A) while the in-plane anisotropy terms $B_{4//}$ and $B_{2//}$ are affine functions of  $A_{2\varepsilon}$.  One can see that a larger precession amplitude $|\delta \theta|_{max}$ first requires a large uniaxial anisotropy, and large in-plane anisotropies $B_{4//}$ and $B_{2//}$; but this will tend to increase precession frequencies high above typical SAW frequencies for micron-sized IDTs. The applied field amplitude and angle can however also be optimized as we show in the following numerical calculations of $|\delta \theta|_{max}$ at fixed strain amplitude $\varepsilon_{max}$=10$^{-5}$  (Fig. \ref{fig:SmallAnglePrecession}). This value is taken deliberately  small to remain in the small perturbation regime.

At fixed field amplitude $\mu_{0}H_{ext}$=24~mT (large enough to insure $\phi_{0}\approx \phi_{H}$, $\theta_{0}$=45$^{\circ}$), the angle of the field is first varied in the plane (Fig. \ref{fig:SmallAnglePrecession}a). The precession amplitude is largest in the $\phi_{H}$=0-90$^{\circ}$ range, with a maximum at  $\phi_{H}$=45$^{\circ}$. This results from the competition between the two in-plane anisotropies terms maximized at $\phi_{0}$=0$^{\circ}$ modulo $90^{\circ}$ (for $B_{4//}>0$), or at $\phi_{0}$=45$^{\circ}$ (for $B_{2//}<0$). The amplitude variations are however weak, and this is clearly not the most critical parameter. At fixed field angle $\phi_{H}$=0$^{\circ}$ this time, the amplitude of the field (and therefore of the initial tilt $\theta_{0}$) is made to vary (Fig. \ref{fig:SmallAnglePrecession}b). The variations observed are this time more pronounced, and $|\delta \theta|_{max}$ is clearly maximum  when the magnetization is most pulled away from its zero-field orientation, as was also concluded from theoretical studies of precession triggered by pico-second acoustic pulses\cite{Linnik2011}. For fields above 30~mT, the magnetization is saturated in the plane, and the precession amplitude plummets down to a few 10$^{-4}$ rad, and gradually decreases to zero. The precise variation of precession amplitude with field amplitude and orientation of course depends on the value of the anisotropy parameters, but since $\omega_{P}$, $\beta$, and $\Gamma$ vary slowly with these parameters, the analytical dependence of $\Omega_{\theta,0}$ with $\phi_{H}$ and $\mu_{0}H_{ext}$ given in Eq. \eqref{eq:OmegaTheta0GaMnAs} gives a good idea of the conditions maximizing this amplitude.

Finally, note that we have not taken into account the influence of the ferromagnetic resonance upon the acoustic wave propagation. We have indeed assumed the phonon-magnon coupling in (Ga,Mn)(As,P) sufficiently weak to neglect in first approximation the absorption of the acoustic wave upon interaction with the ferromagnetic layer. Please refer to Ref. \onlinecite{Dreher12} for a complete analytical treatment of this so-called "back-action" effect.

In summary, analytically solving the LLG equation in the presence of a SAW of given frequency has allowed us to identify the experimental parameters apt to yield the largest precession amplitude in a perturbative regime: field as large as possible without saturating the layer, and applied between [100] and [010] axes. 

\section{Numerical solution: irreversible precessional switching}
\label{sec:AnalyticalSolutionSmallAnglePrecession}

 \setcounter{subsubsection}{0} 
 
\subsubsection{Conditions for precessional switching}

To explore the conditions for precessional switching, it is necessary to go beyond the small angle approximation and solve  Eq. \eqref{eq:LLGa} numerically. The same sample as above is considered, with the field applied in the plane of the layer, along $\phi_{H}$=0$^{\circ}$.

\label{sec:ConditionsForIrreversibleSwitching}

\begin{figure*}
	\centering
		\includegraphics[width=1\textwidth]{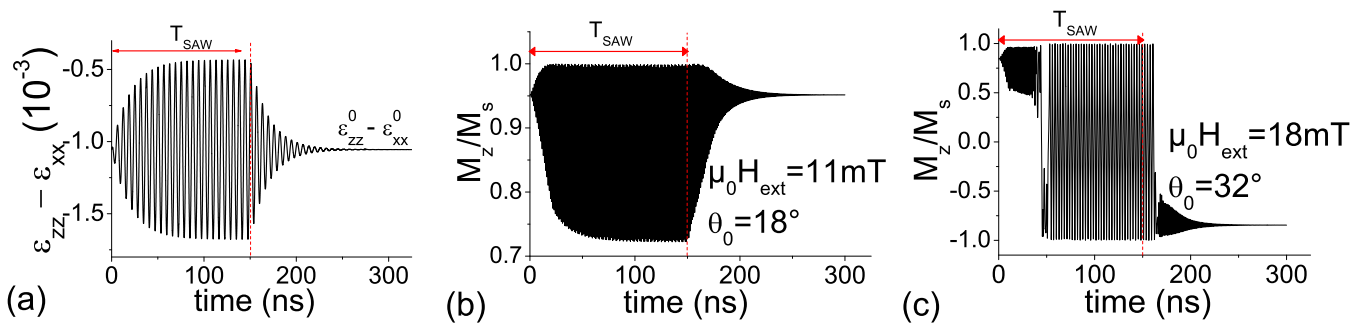}
	\caption{Time behavior of the magnetization from numerical simulations for $\varepsilon_{max}$=2.5.10$^{-4}$, $\alpha$=0.1, $\delta f/f_{P}$=5$\%$ and an initial magnetization pointing towards the upper half, $M_{z}/M_{s}\approx$ 1.  The field is applied along [100]. (a)  Temporal profile of $\varepsilon$(x=0,t) for an rf burst T$_{SAW}$~=~150ns at f$_{SAW}$~=~700~MHz, and a rise time $\tau$=20ns. (b) Large angle precession. (c)  Sustained switching leading to an irreversible reversal at the end of the SAW excitation. The SAW excitation time is indicated by the red dashed line.}
	\label{fig:TroisComportements}
\end{figure*}

Experimentally, SAWs can be excited by rf bursts of length T$_{SAW}\approx$~150ns\cite{Duquesne2012}. Here, we will only consider what happens during a single period. The rise-time $\tau$ of the signal is  given directly by the transit time of the acoustic wave through the emitting IDT. For about 10 pairs of teeth working at sub-GHz frequencies, a realistic value is $\tau~ \approx$~20~ns. The experimental time profile of  $ \varepsilon (x,t)$=$\Delta \varepsilon_{0}$+$\delta \varepsilon (x,t)$ taking into account a rise and decay time (linear experimentally, but modeled as exponential) is computed as shown in Fig. \ref{fig:TroisComportements}a. The SAW's line-width is roughly given by 1/$T_{SAW}$= 7 MHz. Four  main parameters can then be adjusted numerically to explore the different behaviors of the system: the SAW amplitude, $\varepsilon_{max}$ (5.10$^{-5}$-10$^{-3}$), the in-plane magnetic field amplitude, which in turn controls the initial tilt of the magnetization $\theta_{0}$, the detuning of the SAW frequency to the precession frequency  $\frac{|f-f_{P}|}{f_{P}}$, and finally, the damping parameter $\alpha$. Of those parameters, the first two can easily be changed during an experiment. Note that the detuning can equally be defined with respect to $f_{res}$, as they are within 1$\%$ of each other.

At fixed SAW amplitude and damping, two distinct behaviors are observed. Examples are shown in Figs. \ref{fig:TroisComportements}b,c, where the  magnetization initially points upwards ($M_{z}/M_{s}$=1) before the application of the field, and the precession frequency lies around $f_{P}\approx$780-930~MHz for the fields investigated. At low field (11~mT), such that the initial magnetization is moderately tilted towards the plane ($\theta_{0}$=18$^{\circ}$, Fig. \ref{fig:TroisComportements}b) the magnetization remains pointing up during the excitation, and precesses at f$_{SAW}$ in a cone that is widest when the SAW has reached its stationary regime. At the SAW extinction, the magnetization returns to its initial position. This regime will  defined as "large angle precession". Indeed, the amplitude of this precession is about 10-100 times larger than the one observed in picosecond acoustics-triggered precession\cite{Scherbakov2010}: the strain pulse amplitude at the precession frequency  is weak in this latter technique. At larger applied field (18~mT), the magnetization first precesses in the upper quadrant at f$_{SAW}$, fully switches to $M_{z}/M_{s}\approx$~-1, and then oscillates between up and down positions at half the excitation frequency (Fig. \ref{fig:TroisComportements}c). By adjusting  T$_{SAW}$ in the range [$2n/f_{SAW}$,$2(n+1)/f_{SAW}$], the magnetization can then be released in the downwards position. This regime will be named "sustained switching, conditional reversal". It relies on the same mechanism as the precessional switching using tailored ultra-fast magnetic field pulse which was used in garnets\cite{Gerrits2002,Bauer2000} but the lower precession frequencies make the adjustment of  T$_{SAW}$ less constraining. The optimal value of T$_{SAW}$ for reversal is quite experiment-dependent though, as it depends on both the decay time of the SAW (related to the number of teeth in the IDT), and to the magnetization damping $\alpha$.

In between those two behaviors is a transition regime  (not shown) where the magnetization  undergoes quite a chaotic behavior, at times switching irreversibly before the end of the SAW. Finally, for an applied field saturating the magnetization in the plane of the sample ($\theta_{0}$=90$^{\circ}$, not shown), the SAW induces a precession of  $\vec{M}$ around the applied field, with an amplitude that decreases when the strength of the applied field increases.

\begin{figure}
	\centering
		\includegraphics[width=0.485\textwidth]{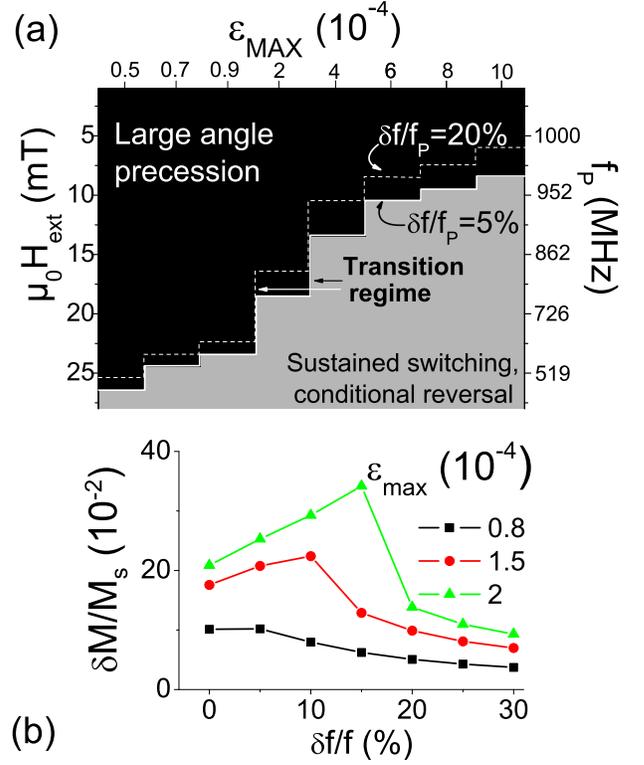}
	\caption{(a) Switching diagram of a macrospin  excited by a SAW with $\alpha$=0.1, a SAW frequency detuned by 5$\%$  from the precession frequency $f_{P}$, and an in-plane field along [100]. The large angle precession regime is separated from the sustained switching regime by the continuous white line (transition regime). Critical fields for  a frequency detuning of 20$\%$ are indicated by the dashed line. (b) Amplitude of the precession  $\frac{\delta M}{M_{s}}$(Fig. \ref{fig:TroisComportements}b behavior) as a function of the detuning, for different strain amplitudes.}
	\label{fig:PhaseDiagram}
\end{figure}

A more thorough exploration of the parameter space is shown in Fig. \ref{fig:PhaseDiagram}a. The SAW amplitude $\varepsilon_{max}$ is varied in steps of  2.10$^{-4}$ or 2.10$^{-5}$,  the in-plane applied field in steps of 1~mT with $\phi_{H}$=0$^{\circ}$, and the frequency detuning is first fixed to 5$\%$. The resulting diagram shows that the behavior of Fig. \ref{fig:TroisComportements} is very generic, regardless of the SAW amplitude: a large amplitude precession regime at low fields (in black), a conditionnal switching regime at high fields (in gray), and in between a transition regime (white line). As expected intuitively when $\varepsilon_{max}$ decreases, the field necessary to obtain switching increases in order to keep the precession wide and compensate for this lesser efficiency. Note that for large strain amplitudes ($\varepsilon_{max}>$5.10$^{-4}$), this generic behavior is maintained, but multiple frequencies appear in the sustained switching regime due to strong non-linearities. The critical fields obtained for a larger detuning (20$\%$) are indicated by the dashed line. Quite counter-intuitively, these seem to be \textit{lower} than the ones found for $\delta$f/f=5$\%$, in particular at higher strain amplitudes, as if a \textit{larger} precession amplitude were obtained away from resonance as opposed to at resonance.  To elucidate this, we set the field to 11~mT and systematically recorded the precession amplitude as a function of the frequency detuning, for three different strain amplitudes. The result is shown in Fig. \ref{fig:PhaseDiagram}b, and confirms that whereas at low strain amplitude, the maximum precession amplitude is indeed obtained at resonance ($\delta$f/f=0$\%$), when the strain amplitude increases,  the maximum precession amplitude can be obtained quite far from resonance, at an increasingly large  frequency detuning. This confirms that the small-perturbations approach of Section III is only valid for small strain amplitudes, and that beyond, the behavior becomes highly non-linear and the system's eigenfrequency is very probably not given by $f_{P}$ anymore, but by a lower frequency. Finally, the main effect of decreasing the damping (not shown) is to \textit{lower} the critical field between large angle precession and precessional switching. The important conclusion of these simulations is that there is a large region of the ($\varepsilon_{max}$, $\mu _{0}H_{ext}$) parameter space where irreversible switching of a macrospin is possible, and this at fairly low fields. 

Note that a more elegant way of identifying switching conditions would be to determine an analytical criterion leading to the growth of non-linearities in the system. This is not trivial in this coupled  $\dot{\theta}$=$f(\phi,\theta,\dot{\theta})$, $\dot{\phi}$=$g(\theta,\phi,\dot{\phi})$ system, and is beyond the scope of this paper.

\subsubsection{Particular cases: i) SAWs propagating along (110),  ii) buried layers}
\label{sec:particularCasesBuriedLayersInfluenceOfTheEpsilonXzComponentAndSAWPropagatingAlong110}

Here we address the case of a Rayleigh wave propagating along a (110) direction, a configuration easier to implement experimentally. A straightforward calculation using a $\pi/4$ rotated frame shows that the SAW once again has three components, denoted $\varepsilon_{XX}$, $\varepsilon_{ZZ}$ and $\varepsilon_{XZ}$, of identical shape and similar amplitude as the one traveling along [100] (details in Annex B). Given this strain tensor, we rotate it back into the x//[100] frame and inject the resulting components in the energy. The free energy density of the layer  is then identical to Eq. \eqref{eq:Fm} with  $ \varepsilon(x,t)$= $\Delta \varepsilon_{0}+\varepsilon_{zz}(x,t)-\varepsilon_{xx}(x,t)$  replaced by: $\Delta \varepsilon_{0}+\varepsilon_{ZZ}(x,t)-\frac{1}{2}\varepsilon_{XX}(x,t)$, and $\varepsilon_{xy,0}$ replaced by $\varepsilon_{xy,0}\pm\frac{1}{2}\varepsilon_{XX}$ (+ for $\vec{q}$//[110], - for $\vec{q}$//[1$\bar{1}$0]).  The calculation therefore requires knowing $\varepsilon_{xy,0}$ which is problematic: there is no experimental measurement of this shear strain, and therefore no independent determination of $A_{2xy}$ and $\varepsilon_{xy,0}$ is possible. We will therefore stick to the $\varepsilon_{xy,0}$=10$^{-4}$ value used in Ref. \onlinecite{Linnik2011}, and deduce from our experimental value B$_{2//}$~=~-1.2~mT the parameter $A_{2xy}$=-12~T\cite{note1_acoustic_2012}. 

The  amplitude of the precession triggered by the SAW is then estimated numerically using the same micromagnetic  parameters as above. For  strain amplitudes $\varepsilon_{max}$=0.5 to 10.10$^{-4}$, and $\mu_{0}H_{ext}$=5~mT along [100], the precession amplitude  when the SAW is traveling along [110] or [1$\bar{1}$0] is systematically about half of when it travels along [100]. This is mainly due to the fact that the strain amplitude in front of the $A_{2\varepsilon}$ coefficient is reduced: the biaxial strain is along the crystallographic axes (100), and effectively reduced along the diagonals (110). The magneto-strictive process is then less efficient.

\vspace{10 mm}

Finally, we look at what happens for either a thick magnetic layer, or a buried layer, since the three Rayleigh wave strain components oscillate in amplitude with  $z$ (Fig. \ref{fig:StrainWave}b). Away from the surface,  $\varepsilon_{zz}$ and $\varepsilon_{xx}$ both reduce to their first zero within about 0.1-0.2$\Lambda_{SAW}$, whilst $\varepsilon_{xz}$ takes over, albeit with a much smaller amplitude. For layers of this order of thickness ($<$300~nm in our case), the conclusions drawn above remain valid. However, an interesting  case is that of a thin magnetic layer buried where $\varepsilon_{xz}$ is maximum, and $\delta \varepsilon$ non-zero. As evidenced in Fig. \ref{fig:StrainWave}b, this lies around $z_{c}$=0.25$\Lambda_{SAW}$, where $\delta \varepsilon$ is then still at about half its surface value.

In the (Ga,Mn)(As,P) free energy density form (Eq \eqref{eq:Gm}), there is no dependence on $\varepsilon_{xz}$ as this strain component is not present statically in the layers. However, just as a $\varepsilon_{xy}$ term will give a magnetic anisotropy between [110] and [1$\overline{1}$0] axes (term $\frac{1}{2}A_{2xy}\varepsilon_{xy}\sin^{2}\theta \sin 2\phi$ in Eq. \eqref{eq:Fm}), it is easy to imagine that a $\varepsilon_{xz}$ term will give a  magnetic anisotropy between [110] and [001] axes,  expressed as: $\frac{1}{2}A_{2xz}\varepsilon_{xz}\sin 2\theta \cos \phi$. This term was evaluated numerically to $A_{2xz}$=80~T (details in Annex C) close to 95~K.

The  amplitude of the magnetization precession triggered by the SAW was then compared at fixed applied field  5~mT, and SAW frequency 2~GHz (5$\%$ detuned from the precession frequencies that are larger than at 95~K since the anisotropy is stronger) and strain amplitude $\varepsilon_{max}$=10$^{-5}$ for a surface layer ($z$=0) or for a  layer buried at $z_{c}$=0.25$\Lambda_{SAW}$. In the latter case, the precession amplitude is about three times larger compared to a surface layer. This is due to the fact that $A_{2xz}$ is large and that the three strain components now contribute to the precession. However, for the precession frequencies explored at 95~K, $\Lambda_{SAW}$ is too large (several microns), and $z_{c}$ therefore unrealistically deep. This option might however prove more relevant for materials requiring higher SAW frequencies (smaller $\Lambda_{SAW}$).


\section{SAW assisted domain  nucleation}
\label{sec:SAWAssistedDomainWallNucleation}

Let us mention briefly another approach for irreversible switching induced by a SAW. More likely than the (coherent) precessional switching of a structure, is the possibility that the SAW will locally lower  a domain nucleation  barrier and thus switch the whole layer under a small (propagating) magnetic field applied concurrently  opposite to the initial magnetization. Indeed, in perpendicularly magnetized layers, the coercive field $\mu_{0}H_{c}$ is largely determined by DW nucleation and/or propagation  barriers\cite{Lacheisserie2004}. Transient reduction of coercivity has already been demonstrated in  garnets\cite{Hohlfeld2001} or magnetic semiconductors\cite{Reid2010,Hall2008} using  ultra-fast light pulses. It is also at the basis of thermo-magnetic writing. 

Neglecting the stray field energy between the nucleated domain and the rest of the layer\cite{Hubert}, the domain nucleation barrier can be expressed as $E_{nuc}$=$2\pi r_{nuc}d \times \sigma $, where  $d$ is the  layer thickness, $r_{nuc}$   the radius of the nucleated domain, and   $\sigma$  its surface energy. The latter depends on the perpendicular uniaxial anisotropy and the exchange constant $A_{ex}$ through $\sigma \propto \sqrt{A_{ex} B_{\bot}(\varepsilon_{xx},\varepsilon_{zz})}$\cite{Hubert}. If the magnetization reversal is nucleation  limited, switching then occurs on a typical timescale given by the Arrhenius law $\tau$=$\tau_{0}\exp (\frac{E_{Z}+E_{nuc}}{k_{B}T})$, where $k_{B}T$ is the thermal energy, and $E_{Z}$=-$\mu_{0}H_{ext}M_{s}\times d\pi r^{2}_{c}$ is the Zeeman energy which lowers this barrier, with $H_{ext}$ the field applied perpendicular to the layer. $\tau_{0}$ is the typical time needed for an energy exchange between spin and lattice, and depends on the damping, the anisotropy constants and the magnetization\cite{Back1999}. It is usually estimated to about 10~ps.

For the sample at 95~K considered above, the effective uniaxial anisotropy field  is $B_{\bot}$$\approx$$ B_{2\bot}$=23~mT. Previous temperature dependent experiments on (Ga,Mn,As,P)\cite{Haghgoo2010} have moreover given $A_{ex}$=10$^{-13}$~pJ/m, providing  an estimation of the DW surface energy $\sigma$=2.10$^{-5}$~J/m$^{2}$. Experimental hysteresis loops done at 90~K yield $\mu_{0}H_{c}$=2~mT, with the perpendicular field applied in $\approx$100~ms pulses. Using the Arrhenius law, this gives an \textit{experimental} estimation of the nucleated domain's diameter   of a few nanometers. The nucleation barrier is then of the order of 44~meV (much lower than the intrinsic 1~eV barrier estimated using DW nucleation theory\cite{Gourdon2007}),  the Zeeman lowering around  -3.10$^{-2}H_{ext}$~meV/mT, while the thermal energy lies around 8~meV. In the presence of the SAW, the effective anisotropy coefficient becomes: $B^{SAW}_{\bot}(x,t)$=$(A_{2\varepsilon}-2A_{4\varepsilon})(\Delta \varepsilon_{0}+\delta \varepsilon (x,t))$. Because the lattice mismatch is  small in this sample (in order to have a weak anisotropy), the transient strain modification $\delta \varepsilon (x,t)$  can very well be of the order of the static strain mismatch $\Delta \varepsilon_{0}$ (Fig. \ref{fig:TroisComportements}a), thus strongly reducing the uniaxial anisotropy field. This dramatically lowers the DW nucleation barrier during about a quarter of the SAW's period. If  the resulting switching time $\tau$ is shorter than the time during which this barrier is very low, ultra-fast magnetization switching is expected to occur under a small (propagating) perpendicular field. Note that this is a non-resonant process, and as such does not involve any strong constraint on the SAW frequency. On the contrary, it might be more relevant to aim for low frequency, which will leave plenty of time for the domain to nucleate during the transient decrease of anisotropy.

\section{Discussion}
\label{sec:Discussion}

\label{sec:In real life}

In the first approach (Sections. III,IV), the calculations and simulations were done for a macrospin at a fixed distance from the combs. For an extended, single-domain structure (no exchange energy), the conclusions will remain identical as long as the shape anisotropy of the structure is not significantly different from the $\frac{\mu_{0}M_{s}}{2}\cos^{2}\theta$ term of Eqs. \eqref{eq:Fm},\eqref{eq:Gm}. For micron-sized structures, and given the weak magnetization of (Ga,Mn)(As,P), this is  legitimate. However, for samples with a strong shape anisotropy, a modified demagnetization factor would need to be included in the energy form, as was done by Roy et. al\cite{Roy2011a}. A more proper solution would however eventually need to take into account the exchange contribution in the free energy. In both picosecond acoustic pulses- and SAW-induced precession, the modification of the eigen-frequencies due to this exchange contribution was shown to be negligible\cite{Bombeck2012,Dreher12}, given the dispersion curve in the considered regime remains almost flat. An estimation of this contribution to the real space, i.e the generation of spin waves to accommodate a spatially varying magnetization vector is a problem more appropriately tackled by micromagnetic finite elements methods, and is beyond the scope of this paper.

In the second approach,  a phenomenological model of SAW-assisted domain nucleation is used where exchange is implicitly taken into account (formation of a domain wall), and the field applied perpendicularly to the plane. Once again, SAWs seem more adequate than picosecond acoustic pulses. Indeed, were the transient strain  strong enough at the relevant frequency to induce a substantial change of the uniaxial anisotropy coefficient to lower the domain's nucleation barrier, this would only last about 10~ps\cite{Thevenard2010a}, which would require an ultra-fast DW nucleation. Looking at the Arrhenius law once again shows that this would mean effectively canceling the energy barrier, in order to have $\tau$$\approx$$\tau_{0}$, which seems unlikely given the weak accessible transient strain amplitudes in this technique.

\section{Conclusion}
\label{sec:Conclusion}

A general analytical approach to SAW-assisted magnetization precessional switching has been developed based on the strain dependence of the magnetic anisotropy coefficients, and taking the damping into account. Several parameters were found to be important, such as a large in-plane field, and SAWs propagating along [100] rather than [110]. Numerical simulations using realistic experimental parameters of (Ga,Mn)(As,P) then clearly evidenced a wide range of fields and SAW amplitudes under which irreversible switching was possible. SAWs were shown to possibly be a more adequate method to switch magnetization than picosecond acoustic pulses. Finally, although these concepts were tested on (Ga,Mn)(As,P) thin films, the analytical forms of energy and precession amplitude given in this work make them applicable to any magneto-strictive material.


\label{sec:ACKNOWLEDGEMENTS}

We gratefully acknowledge insightful advice from C. Tanguy at Orange labs, and technical helop from S. Majrab and L. Becerra. This work was performed in the framework of the MANGAS project (ANR 2010-BLANC-0424-02).

\section*{Annex}

\subsection{Magnetic anisotropy coefficients}
\label{sec:MagneticAnisotropyCoefficients}

The magnetic anisotropy terms  $A_{4\varepsilon}$, $A_{2\varepsilon}$, $A_{2xy}$, and $B_{c}$ of Eq. \eqref{eq:Fm}, and the terms $B_{2\bot}$, $B_{4\bot}$, $B_{2//}$ and $B_{4//}$ (obtained experimentally by ferromagnetic resonance experiments for instance) of Eq. \eqref{eq:Gm} are related as follows:

     \begin{eqnarray*}
A_{4\varepsilon} 		   &=&	\frac{B_{4//}-B_{4\bot}}{6\Delta \varepsilon_{0}}\\
A_{2\varepsilon}		     &=&	\frac{B_{4//}-B_{4\bot}}{3\Delta \varepsilon_{0}}+\frac{B_{2//}-2B_{2\bot}}{2\Delta \varepsilon_{0}}\\
B_{c}         		   &=& -\frac{B_{4\bot}+2B_{4//}}{6}\\
A_{2xy}              &=&  \frac{B_{2//}}{\varepsilon_{xy,0}}\\
where\  \Delta \varepsilon_{0}  &=& \varepsilon_{zz,0}-\varepsilon_{xx,0}\\
     \end{eqnarray*}

For the vast majority of (Ga,Mn)(As,P) samples, the relationship $A_{4\varepsilon}<<A_{2\varepsilon}$ holds. 

Note that there is no experimental evidence of an $\varepsilon_{xy,0}$ shear strain, but rather it is a physical effect of the same symmetry that is at the root of the weak uniaxial anisotropy  $B_{2//}$. During the growth, when atoms are mobile on the surface, nearest-neighbor Mn pairs on the GaAs (001) surface have a lower energy for the [1-10] direction compared to the [110]\cite{Birowska2012}.

  \subsection{Derivation of the strain wave expression}
\label{sec:StrainWave}

The strain components of the Rayleigh wave propagating in a cubic material along $\vec{q}$//[100] are quite different from the usual formulas found in textbooks for isotropic materials [Ref JYQ]. They can be found analytically as:

    \begin{eqnarray*}
         \label{eq:epsxx}
  \varepsilon_{xx}(r,z,t) & = & -2i \zeta_{0}q ~e^{i \psi_{2} / 2}~ e^{-a q z} \cos(b q z + \psi_{2}/2)e^{i(\omega t-\vec{q}.\vec{r})} \\
         \label{eq:epszz}
  \varepsilon_{zz}(r,z,t)& = & -2~i~\rho~ \zeta_{0}qe^{i\psi_{2} /2} e^{-a q z}[-a \sin(b q z  + \psi_{2}/2 -\psi_{1})+\\
                      &   & b \cos(b q z  + \psi_{2}/2-\psi_{1} )]e^{i(\omega t-\vec{q}.\vec{r})}\\
         \label{eq:epsxz}
  \varepsilon_{xz}(r,z,t) & = &  -   \zeta_{0}qe^{i \psi_{2} / 2}~ e^{-a q z}[ a \cos(b q z  + \psi_{2}/2) +\\  
  										&   &b \sin(b q z + \psi_{2}/2) + \rho \sin(b q z + \psi_{2}/2- \psi_{1})]e^{i(\omega t-\vec{q}.\vec{r})}
     \end{eqnarray*}

We define  the wave-vector $\vec{q}$ (norm $q$),    the position $\vec{r}$ (norm $r$) and $\zeta_{0}$ the amplitude of the displacement. When $\vec{q}$//(100), $\vec{q}.\vec{r}$=$qx$ and for a magnetic layer much thinner than $\Lambda_{SAW}$, we can set $z$=0 (Fig. \ref{fig:StrainWave}), and simplify $\delta \varepsilon (x,t)$=$\varepsilon_{zz}(x,0,t)$-$\varepsilon_{xx}(x,0,t)$ into: $\delta \varepsilon (x,t)$=$\varepsilon_{max}\cos(\omega t-qx)$, where $\varepsilon_{max}$ is the amplitude of the resulting wave.

 The parameters are related by: $a \cos \psi_{2}/2+b \sin \psi_{2}/2=\rho\sin (\psi_{1}-\psi_{2}/2)$. For GaAs and $\vec{q}$//(100), the values found numerically are:  $\psi_{1}$=-0.328, $\psi_{2}$=-1.9,  $\rho$=1.18, $a $=0.402 and $b$=-0.561. For $\vec{q}$//(110), these coefficients are only slightly modified:  $\psi_{1}$=-0.531, $\psi_{2}$=-2.1,  $\rho$=1.34, $a$=0.500 and $b$=-0.480, yielding strain amplitudes about 15$\%$ smaller than along (100) directions.


\subsection{Effect of a $\varepsilon_{xz}$ strain component on magnetization precession}
\label{sec:EffectOfAEpsilonXzStrainComponentsOnMagnetizationPrecession}

In the presence of a $\varepsilon_{xz}$ strain component, one can expect a magneto-strictive component\cite{Linnik2011} of the form $A_{2xz}\varepsilon_{xz}m_{x}m_{z}$=$\frac{1}{2}A_{2xz}\varepsilon_{xz}\sin 2\theta \cos \phi$.  This term can be evaluated theoretically using an effective mass Hamiltonian with the six-band k.p Luttinger-Kohn term, a strain tensor,
and  the p-d exchange interaction of the holes and the Mn spins in the molecular-field approximation\cite{Zemen2009,Dietl2001}. The saturation magnetization was set to 6~kA.m$^{-1}$ ($T\approx$~95~K), which corresponds to a $|B_{G}|\approx$4~meV spin splitting parameter\cite{Dietl2001}. The usual biaxial strain terms $\varepsilon_{zz,0},\varepsilon_{xx,0}$ were set to zero, and a non-zero term $id\varepsilon_{xz}$ introduced in the Bir-Pikus strain tensor, where $d$=-4.8~eV is the shear deformation potential\cite{Zemen2009}. In this way, $B_{c}$ and $A_{2xz}$ were the only unknown parameters in the free energy density. The energy difference $F(\theta)-F([001])=Bc(\cos^{4} \theta+\sin^{4} \theta)+\frac{1}{2}A_{2xz}\varepsilon_{xz}\sin 2\theta$ was then computed and fit numerically for $p$=3.10$^{20}$~cm$^{-3}$ yielding:  $A_{2xz}$=80~T. This value is quite large, in fact larger than any of the anisotropy parameters, but the resulting anisotropy field expected to be less than 10~mT.

\subsection{Small angle magnetization precession amplitude}
\label{sec:FullExpressionOfTheSmallAngleMagnetizationPrecessionAmplitude}

The precession amplitude $\delta \theta$ given in Eq. \eqref{eq:deltatheta0} depends on various parameters given below:

\begin{widetext} 
		\begin{align}
 \label{eq:ftheta}
f_{\theta}  &= -\frac{\gamma}{\sin \theta_{0}}F_{\phi \varepsilon_{zz}}\\
	\label{eq:fphi}
f_{\phi}  &= \frac{\gamma}{\sin \theta_{0}}F_{\theta \varepsilon_{zz}}\\
	\label{eq:f}
f(\omega,\beta)&=\sqrt{\frac{(\omega \beta)^{2}+(\omega_{P}+\chi \beta)^2}{1+\beta^2}}\\
	\label{eq:beta}
\tan \beta&=-\frac{\omega_{P}}{\gamma}\frac{2(1+\alpha^{2})}{\alpha H_{\alpha}+\frac{2(1+\alpha^{2})}{\sin \theta_{0}}\left[\frac{f_{\theta}F_{\theta \phi}+f_{\phi}F_{\phi \phi}}{f_{\phi}\alpha \sin \theta_{0}-f_{\theta}}\right]}\\
	\label{eq:omegatheta}
\Omega_{\theta}&=-\sqrt{ \frac{[f_{\phi} \alpha  \sin \theta_{0}-f_{\theta }]^2+[f_{\phi}\sin \theta_{0}(2F_{\phi \phi}+\alpha^2(F_{\phi \phi}-F_{\theta \theta}\sin^2 \theta_{0}))+f_{\theta}(\alpha H_{\alpha}\sin^{2} \theta_{0}+2F_{\theta \phi}\sin \theta_{0}(1+\alpha^{2}))]^2}{\sin^4 \theta_{0}[(1+\alpha^2)\frac{4\omega^{2}_{0}}{\gamma^{2}}-\alpha^{2}H^{2}_{\alpha}]}}
		 \end{align}		
\end{widetext}

In the case of (Ga,Mn)(As,P), some approximations can be made using $A_{4\varepsilon}<<A_{2\varepsilon}$ and developing $\Omega_{\theta}$ to its zero-th order expansion in $\alpha$:\\

		\begin{align}
f_{\theta}  &=  -\gamma A_{4\varepsilon}\sin^{3} \theta_{0}\sin 4\phi_{0}\\
	\label{eq:fphi}
f_{\phi}  & \approx  -2 \gamma A_{2\varepsilon} \cos \theta_{0}\\
 \label{eq:OmegaTheta0}
\Omega_{\theta,0}&=-\sqrt{f^{2}_{\theta}+\frac{(f_{\theta} F_{\theta \phi}+f_{\phi} F_{\phi \phi})^{2}\gamma^{2}}{\sin^{2}\theta_{0}\omega^{2}_{0}}}
		 \end{align}

\bibliographystyle{phjcp}

\end{document}